# Regression discontinuity design in perinatal epidemiology and birth cohort research

Popovic M[1], Zugna D[1], Richiardi L[1]

**Affiliations**

[1] Cancer Epidemiology Unit, Department of Medical Sciences, University of Turin, Turin, Italy

**Corresponding author**

Maja Popovic

Cancer Epidemiology Unit, Department of Medical Sciences, University of Turin, Italy

E-mail: maja.popovic@unito,it

**Key words:** Regression discontinuity, life course epidemiology, study design, review, natural experiment, children, pediatric, perinatal

**Funding:** This work was funded by the European Union's Horizon 2020 research and innovation programme (Grant Agreement No. 733206 LifeCycle). This manuscript reflects only the author's view and the Commission is not responsible for any use that may be made of the information it contains

**Conflicts of interest:** none to report

**Working paper: August 2022**

**Abstract**

Regression discontinuity design (RDD) is a quasi-experimental approach to study the causal effects of an intervention/treatment on later health outcomes. It exploits a continuously measured assignment variable with a clearly defined cut-off above or below which the population is at least partially assigned to the intervention/treatment. We describe the RDD and outline the applications of RDD in the context of perinatal epidemiology and birth cohort research.

There is an increasing number of studies using RDD in perinatal and pediatric epidemiology. Most of these studies were conducted in the context of education, social and welfare policies, healthcare organization, insurance, and preventive programs. Additional thematic fields include clinically relevant research questions, shock events, social and environmental factors, and changes in guidelines. Maternal and perinatal characteristics, such as age, birth weight and gestational age are frequently used assignment variables to study the effects of the type and intensity of neonatal care, health insurance, and supplemental newborn benefits. Different socioeconomic measures have been used to study the effects of social, welfare and cash transfer programs, while age or date of birth served as assignment variables to study the effects of vaccination programs, pregnancy-specific guidelines, maternity and paternity leave policies and introduction of newborn-based welfare programs.

RDD has advantages, including relatively weak and testable assumptions, strong internal validity, intuitive interpretation, and transparent and simple graphical representation. However, its use in birth cohort research is hampered by the rarity of settings outside of policy and program evaluations, low statistical power, limited external validity (geographic- and time-specific settings) and potential contamination by other exposures/interventions.

## 1. Introduction

Since the first Barker's studies in the mid-1980s, focused on the influence of foetal undernutrition and growth during early periods of development on different short- and long-term health outcomes,[1–3] etiologic non-communicable disease research increasingly shifted its attention on early development – a critical window for prevention of later disease. In recent years, a growing number of birth cohort and other longitudinal studies with enormous amount of data collected across different periods of life have become available for life course epidemiology. The selection of appropriate study design for complex research questions, multiple relationships of biological, social, and other contextual variables, repeated measures over time, and residual confounding that threats the estimation of causal effects remain the main methodological challenges in the context of life course epidemiology.[4]

The assumption of no unmeasured confounding is probably the main obstacle to causal inference within the context of non-experimental studies, and several analytical and design approaches have been developed and extensively used in the past decades to control for confounding and obtain a potentially unbiased estimate of the treatment/exposure effect. These include, but are not limited to, multivariable regression analysis, exposure and outcome negative control, within-sibling design, and instrumental variable analysis. Natural experiments, such as epidemics and famines, have been historically used to establish causal relationships and to study the causes of health outcomes.[5,6] A series of studies of the Dutch "Hunger Winter" of 1944 on the health effects originating from the prenatal period,[7–9] were among the first applications of natural experiments in the context of birth cohort research. These studies exploited random or as-if random treatment or intervention assignments of the population that may stem from various natural or pseudo-natural sources. Methods typical applied in these settings involve instrumental variable analysis, difference-in-differences, interrupted time series, and regression discontinuity.[10] However, the relative rarity of "shock events" or strong exogenous instruments for exposures, the need of more complicated research designs, the involvement of small and more gradual effects, the small proportion of population affected by the exposure and the need of prompt data collection prevent the widespread use of natural experiments in life course epidemiology. Studies exploiting the geographical or temporal variations in medical prescriptions or population-level policies are more common. These studies use the "quasi-naturally" occurring variation in exposure to identify the effect of an event/exposure on an outcome of interest.

Here, we focus on regression discontinuity design (RDD), a quasi-experimental approach, applied in circumstances where an exogenous source of variation arises from a continuously measured assignment variable with a clearly defined cut-off point above or below which the population is at least partially assigned to a treatment or exposure. The assignment variable creates a discontinuity in the probability of the exposure at the threshold, where the direction and magnitude of the jump is a direct measure of the causal effect of the exposure on the outcome for subjects near the cut-point. This approach has been widely applied in the context of natural experiments and by the end of the 20[th] century the RDD has become popular in the econometrics, educational and social research exploiting the threshold rules often used by educational institutions, public and private insurance schemes, governmental welfare programs and social policies.[11,12] Although in many of these studies the main outcomes of interest were health outcomes, the examined program and policy interventions led to their publications mostly in the economics journals and their findings caused generally less attention in the health and epidemiology literature.[12–14]

A growing number of birth cohorts and multiple birth cohort consortia are being established to study the effects of early life exposures on later health outcomes. The RDD is little used in this context. We thus describe the RDD, highlighting its underlying assumptions, advantages, limitations, and approaches for testing the assumptions and the validity of the design, and focus on the studies, assignment variables and interventions/exposures that have been investigated in the context of perinatal and pediatric epidemiology.

## 2. Regression Discontinuity Design – basic concepts and frameworks

RDD, introduced in 1960s,[15] is a quasi-experimental design that shares similarities with randomized controlled trials, but lacks the completely random assignment to the intervention (intervention, treatment, or exposure, hereafter referred as exposure in general). This type of study design typically implies that a researcher or whoever imposes a certain policy, program, or clinical decision control the assignment to the exposure using an a priori, often administratively, decided criterion (e.g., an eligibility rule, or a clinical decision-making guideline). The exposure assignment in RDD studies is thus based on the cut-off value of a continuously measured variable, the assignment variable (also referred to in the literature as the "forcing", "rating" or the "running" variable), that creates a discontinuity in the probability of the exposure at the cut-off point. The assignment variable can be any continuous variable measured before the exposure, provided that individuals cannot manipulate the value of this variable to systematically place themselves above or below the cut-off. The stronger is the individuals' inability to control their own value of the assignment variable the more valid is the design to identify the causal effects. As the cut-off value in the assignment variable is typically determined by an administrative decision, clinical guideline, or some "shock event", it is unrelated to the baseline pre-exposure characteristics of the individuals near the cut-off. This implies that individuals just below the known cut-off are on average similar in all observed and unobserved baseline characteristics to those just above the cut-off except for the exposure of interest, i.e., they are exchangeable. The dissimilarity (i.e. lack of exchangeability) between these two groups typically increases far from the cut-off rendering the validity of the design plausible for relatively narrow windows around the assignment variable cut-off. If the assumption of exchangeability holds, any difference in the outcome on the two sides of the cut-off will be caused by the exposure. The magnitude of the discontinuity in the outcome at the cut-off represents thus the average treatment effect around the cut-off point, where exposed and unexposed individuals are most similar.

The RDD design, formalized by a causal diagram – a Direct Acyclic Graph (DAG),[16,17] is represented in *Figure 1*. The panel A of *Figure 1* shows that the assignment variable ($X$) determines the eligibility status ($D$) for an exposure / treatment ($E$) and is a direct cause of an outcome of interest ($Y$). $U$ denotes measured and unmeasured confounding factors. In a standard causal inference approach the assignment variable represents a confounding variable in the path between $E$ and $Y$, which can be controlled for, but which conditioning is not enough to block the backdoor confounding path through other measured and unmeasured variables ($E \leftarrow U \rightarrow Y$). In the close window around the cut-off ($c$), the assignment variable still determines the eligibility status, but it does not directly affect the outcome and is not affected by other observed and unobserved variables. This means that the eligibility becomes independent of other factors and that it can be used as an instrument for the exposure at the cut-off.

The main condition and assumptions of RDD can be therefore summarized as follows:

i)      A continuous pre-exposure variable with a clearly defined cut-off value for the exposure assignment ("Assignment rule condition");

ii)     No sign of manipulation of the "eligibility criteria" to be below or above the assignment variable cut-off;

iii)    Individuals close to the cut-off are exchangeable. This assumption implies also that the outcome probability function is continuous at the cut-off in the absence of the exposure.

The power of RDD lays in the fact that it can provide valid causal estimates under relatively weak assumptions compared to other observational study designs,[11,18,19] and that most of these assumptions can be tested empirically.

(Figure 1 here)

There are two main conceptual and inference frameworks in RDD that involve slightly different estimation procedures and interpretation of the causal effects: the continuity-based approach, and the local randomization approach. In this paper we focus only on the conceptual commonalities of the two frameworks, while the analytical estimation methods of the two approaches are discussed in detail elsewhere.[20,21] Briefly, the main assumption behind both frameworks is that individuals just below the cut-off are comparable to the individuals just above the cut-off in all relevant characteristics, except for their exposure status. The general logic of measuring a difference in regression lines at the cut-off applies to both frameworks but the main differences are the estimation methods and statistical inference. The continuity-based framework, the most widely used method in practice, is based on the continuity of average potential outcomes near the cut-off, and it typically uses polynomial methods to approximate the regression functions on the two sides of the cut-off (least-squares methods to a polynomial of the observed outcome on the assignment variable).[19,20,22–24] For the estimation it is possible to use either all the observations available around the cut-off (parametric methods) or only observations in a selected window around the cut-off (local or non-parametric methods). According to the local randomization framework, instead, RDD is seen as a randomized experiment near the cut-off, which imposes somewhat stronger underlying assumptions than the continuity-based approach (instead of the continuity of the unknown regression functions at the cut-off, one assumes random assignment in a narrow window around the cut-off with the assignment variable being unrelated to the average potential outcomes).[21,25–27] The estimation is then based on standard methods for random experiments, such as finite-sample Fisher's methods, or on the Neyman's approach with large-sample approximations.[21]

## 3. Sharp and fuzzy regression discontinuity design

In the RDD, the exposure is determined by the assignment rule either completely (deterministically) or partially (probabilistically). When the assignment rule perfectly determines the exposure (from 0 to 1 at the cut-off), the regression discontinuity takes a sharp design. This means that the exposure assignment and the actual exposure status coincide, i.e., all individuals above the cut-off are assigned to an exposure and are exposed, while all those below the cut-off are assigned to the unexposed group, with no cross-over effects. If the assignment rule affects the probability of exposure creating a discontinuous change at the threshold, without an extreme 0 to 1 jump, regression discontinuity takes a fuzzy design. In this setting, there are exposed and unexposed individuals both above and below the cut-off, but the probability of being exposed jumps discontinuously at the cut-off.[11,18]

The sharp and fuzzy RDD are shown with hypothetical examples in *Figure 2*. Sharp RDD is generally appropriate in scenarios of imposed policy measures and RDD in time (time as the assignment variable). Panel A shows a simulated sharp RDD where, for example, a family poverty index can be used as an assignment variable, which value of at least 30 is used as an eligibility criteria for governmental conditional cash transfer at childbirth. Given that all families above this cut-off receive the benefit, while those below the cut-off do not receive it, there is a sharp 0-to-1 jump at the cut-off. As we expect that families with a poverty index just below 30 are very similar to those just above the cut-off, except from the conditional cash transfer, we could estimate the effect of conditional cash transfer on later child outcomes around this cut-off.

Panel B in Figure 2 depicts a hypothetical example (simulated data) of a fuzzy RDD motivated by the study of Daysal et al.[28] The authors investigated the effect of the obstetrician supervision of deliveries on the short-term infant health outcomes, using a national rule of 37 gestational weeks (259 days) at delivery for obstetrician's instead of midwife's delivery supervision. In this scenario, the probability of obstetrician supervision does not "jump" from 1 to 0, meaning that not all the deliveries before 37 completed gestational weeks are supervised by an obstetrician and that some deliveries after 37 gestational weeks are however under the care of an obstetrician for reasons other than prematurity, such as complications during delivery and slow delivery progression. Thus, the simulated discontinuity in the probability of the exposure (obstetrician delivery supervision) represents a fuzzy type of the RDD.

(Figure 2 here)

The sharp and the fuzzy RDDs are similar in a way that the identification procedure, i.e., the main condition of the continuity at the cut-off, is identical, and the general logic of evaluating the difference in regression lines at the cut-off applies to both. However, the estimation procedure differs between the two designs due to crossovers at the cut-off or non-compliance present in the fuzzy design. In addition to the abovementioned general RDD assumptions, the estimation of the causal effects in the fuzzy RDD requires some additional assumptions, as discussed below in section 4.

Several modifications to the two general RDD settings have been proposed in literature. In particular, kink RDD is the extension of the sharp or fuzzy design where rather than the jump in the exposure, the cut-off determines the change in the first derivative.[29] The main assumptions of the kink RDD are the same of fuzzy and sharp RDD, i.e., the individuals on the either side of the cut-off are similar in the baseline characteristics, but instead of estimating a jump in the intercept, kink RDD estimates a change in slope at the cut-off. RDD with multiple cut-offs in the assignment variable, and RDD with composite assignment rules dependent on multiple assignment variables have been also addressed in literature.[30–32] A special case of RDD is a RDD in time where the assignment variable is a calendar time and a known date of intervention, policy change or some other type of shock event is used as a cut-off.[33] This type of RDD is very similar to an interrupted time series or a simple pre-post comparison, with the difference that RDD relies on a bandwidth estimator, and is, thus, useful if an effect in a narrow window around the cut-off is meaningful.

**4. Testing the validity and assumptions of regression discontinuity design**

The main assumptions of the RDD and their potential to be tested empirically are summarized in **Table 1**.

(Table 1 here)

*Assignment rule condition*

The main condition of RDD is the existence of a continuous pre-exposure variable with a clearly defined cut-off value for the exposure assignment. This implies that the assignment variable cannot be influenced by the exposure and that it temporally precedes exposure. The assignment rule is an administratively imposed or recommended value of the assignment variable that determines the eligibility for a policy or program implementation, the introduction of an invention or treatment, or a guideline-based clinical decision. This condition can be empirically tested by plotting the relationship between the exposure and the assignment variables. A discontinuous change in the probability of the exposure at the assignment variable cut-off (*Figure 2*) suggests the possibility of using RDD but is not enough to determine its internal validity. In the case of sharp RDD this discontinuity is defined as the jump in the probability of the exposure from 0 to 1 and is relatively easy to visualize. In fuzzy RDD the observed discontinuities in the exposure at the cut-off are often relatively small, and the smaller the jump in the probability of the exposure is, the weaker the validity of the design is.

In addition to the exposure discontinuity at the cut-off, a graphical presentation of the relationship between the exposure and the assignment variables allows the examination of discontinuities in the exposure at locations other than the cut-off. Any such additional discontinuities threat the validity of RDD as they may indicate other exposures or interventions that could confound the estimate of the causal effect of the exposure of interest. Every RDD analysis must have a clear description of the assignment rule, and the underlying assignment process and alternative hypotheses must be excluded by providing evidence (often only theoretical) that the same cut-off value of the assignment variable is not used to assign the individuals to other exposures that could affect the outcome.

In identifying possible RDD settings it is very important not to mistake the discontinuity with the non-linearity. It is possible, that two local linear regressions fitted on the two sides of the cut-off produce spurious jump in the predicted probability of the exposure when the underlying relationship is non-linear.[34] This is true also for the final step of RDD when looking at the discontinuity in the outcome and will be discussed more in detail below. Another issue to have in mind when plotting the exposure probability as a function of the assignment variable is the width of bins in the assignment variable within which the mean exposure probability is calculated. Too narrow bins allow visualizing the underlying data pattern but can at the same time hide or underestimate the magnitude of the jump. On the other hand, wide bins eliminate noise in the data but hide the underlying data distribution. The impact of the bin width on the graphical presentation of the relationship between the exposure and the assignment variables is shown in Figure 3 using the same data of Figure 2b. To avoid subjective selection of the bin width and potential manipulations in graphical presentations, formal tests and data-driven bin width selection are standard procedures in RDD analyses.[11,35]

(Figure 3 here)

*Lack of manipulation in the assignment variable*

After examining the presence of discontinuity in the exposure at the cut-off, it is important to test two fundamental assumptions determining the validity of the design. The first assumption of no manipulation in the assignment variable means that the cut-off value is exogenous – unrelated to the individuals' value of the assignment variable, and at the same time individuals' assignment variable values are not determined by the imposed cut-off in the assignment variable. In practice, when there is a benefit in receiving an exposure/treatment, the manipulation in assignment variable occurs when the treatment assignment rule is public knowledge and individuals just barely qualifying for a desired exposure/treatment manage to cross the cut-off, with few individuals remaining just below the threshold (imagine a cut-off for final high-school grade for entering university and individuals with a bit more effort manage to cross it in the last moment). In a setting like this, individuals in a narrow window below and above the cut-off are no longer similar and the internal validity of the RDD is compromised. On the other hand, even if individuals are unable to precisely manipulate the assignment variable, the variation in treatment/exposure near the cut-off may still not be as good as random.[11] This can happen with administrative procedures that non-randomly affect the position of individuals in the assignment variable near the cut-off rendering the continuity assumption less plausible.[36,37] It is also important to note that the assumption of no precise manipulation in the assignment variable is not an issue in designs where assignment variables are completely exogenous with no possibility of manipulation in either direction, such as RDD in time (e.g., age, calendar year). Thus, one of the most important steps in performing a valid RDD is a complete knowledge of the data generation process underlying the assignment rule and the understanding of the assignment variable's susceptibility to manipulation.

There is a formal test of the control over the assignment variable that complements the knowledge of the exact assignment mechanism. A strategically manipulated position in the assignment variable creates a discontinuity in the distribution of the assignment variable at the cut-off. For example, if a desirable treatment is assigned if individuals are above the cut-off, and individuals can precisely control over the assignment variable, we expect the density of the assignment variable to be close to zero just below the cut-off and positive just above the cut-off. This can be tested both visually using a graphical representation and formally using the McCrary density test.[38] The basic idea behind the McCrary test is that the marginal density of the assignment variable without manipulation should be continuous around the cut-off. This can be visualized graphically with the density plot of the assignment variable, where any discontinuity/jump in the density near the cut-off is suggestive of some degree of manipulation (*Figure 4*, the graphs are generated using simulated data and the STATA rddensity command[39]). The McCrary test is implemented as a Wald test of the null hypothesis that the discontinuity is zero.[38] As pointed out in the original McCrary paper, this test will be informative only when the manipulation of the assignment variable is monotonic, i.e., manipulation shifts individuals in one direction only. Several other modifications and refinements of the assignment variable manipulation test have been proposed in literature and are widely implemented in the existing statistical applications.[39–41]

(Figure 4 here)

*Exchangeability around the assignment variable cut-off*

The exchangeability assumption, also called the continuity assumption, implies that individuals just above and below the cut-off are similar with respect to the distribution of observed and unobserved factors, except from the treatment/exposure, and thus they have the same expected outcome if subject to the same exposure level.[19] This assumption was formalized using the Rubin's potential outcomes framework of causal inference,[42,43] that for each individual imagines a pair of potential outcomes, one for what would occur if the individual were exposed to the treatment/exposure, and one if not exposed. If it were possible to observe both potential outcomes simultaneously, we would be able to estimate the causal effect of the treatment/exposure as the difference between the two potential outcomes at an individual level. As this is possible only in theory, average treatment effects are estimated over populations. The exchangeability/continuity assumption in the RDD setting is, thus, formalized as the continuity of the potential outcomes at the area of the cut-off, which allows us to use the average observed outcomes of individuals just below the cut-off as a counterfactual for those that are just above the cut-off.

As the continuity assumption involves continuous conditional expectation functions of the potential outcomes through the cutoff, and the potential outcomes are unobservable, it is not directly testable. However, there is a way to test its implications. If the continuity assumption holds and individuals immediately below the threshold are a valid counterfactual for those immediately above the threshold, then the distribution of observed baseline characteristics is also expected to have similar distributions in these two groups. In a valid RDD, these distributions tend to be the same as we approach the cut-off (narrower bandwidths around the cut-off). As a result, we can indirectly test the continuity assumption by looking at the distribution of observed baseline covariates that by definition are not influenced by treatment/exposure, and should not be changing at the cut-off. Any discontinuity of the observables at the cut-off indicates a violation in the underlying assumption and calls into question the RDD validity.

In practice, it is advisable to perform both a graphical inspection and a formal testing. The graphical representation is a series of simple plots of the relationship between the baseline covariates not affected by treatment/exposure and the assignment variable. For example, in a hypothetical study of the effect of the type of neonatal care on offspring outcomes where 2500 grams is a birth weight cut-off for intensive neonatal care, we should examine the distribution of the observed maternal baseline characteristics (e.g., age, education level, pre-pregnancy body mass index (BMI), household income) around the 2500 grams of birth weight (*Figure 5,* data from the NINFEA birth cohort[44], which is approved by the local Ethical Committee -approval n. 45, and subsequent amendments). Discontinuity in any of the baseline observables around the cut-off, in our hypothetical example maternal pre-pregnancy BMI, threats the validity of the design.

(Figure 5 here)

As there are several approaches to estimation in RDD (see below for details and further readings), to test the balance of covariates at the cut-off one should follow the same visualization, test statistic, and inference procedures used in the main RD analysis for the outcome of interest. For example, nonparametric local polynomial regression-discontinuity estimation techniques could be applied if the estimation is done by the continuity approach,[20,22–24,45] or methods from the classical analysis of experiments and potential outcomes could be used in the case of local randomization approach.[21,25–27] Independently of the RRD conceptual framework, if there are many covariates to test, some discontinuities may be observed by random chance only. In this setting, the multiple testing problem may become an issue with an over-rejection of the null hypotheses. To control the family-wise error rate or false-discovery rate one may apply the Bonferroni or Benjamini-Hochberg multiple testing corrections, respectively.[46,47] As we are generally not interested in a single covariate test, but in an overall joint null hypothesis of the baseline covariates balance, alternative approaches have been proposed to combine the multiple tests into a single test statistic,[48,49] but these are seldom used in the current RDD literature.

Covariate adjustment has been proposed and broadly used with the aim of improving the validity of the RRD design in situations with imbalances in the distribution of the baseline covariates at the cut-off. It has been, however, shown that the validity of the RDD cannot be fixed by the covariate adjustment and that additional strong assumptions are needed to interpret covariate-adjusted RDD estimates as causal.[50,51] In general, the adjustment for covariates in RDD can be useful to improve efficiency and statistical power, with the treatment point estimate being stable to the adjustment.[50]

*Additional assumptions for fuzzy RDD*

Settings with a sharp RDD are unfortunately not so common in practice. Imperfect exposure/treatment take-up and factors other than the assignment rule that affect the probability of assignment are frequent issues that call for the use of a fuzzy RDD. In a fuzzy RDD, the change in the probability of being treated/exposed at the cut-off is always less than one, and this RDD setting has a close analogy to the noncompliance in randomized controlled trials, where one is interested in both the effect of being assigned to treatment and the effect of receiving treatment on the outcome of interest. The former involves the estimation of the average "intention-to-treat" effect and has the properties of a sharp RDD (perfect compliance to assignment), while the latter requires additional assumptions (imperfect compliance to treatment/exposure take-up). The fuzzy RDD estimates the local average treatment effect (LATE)[52], which is the average treatment effect for the compliers. The estimation is similar to the Wald's treatment effect in an instrumental variable (IV) design (it can be recovered using a two-stage least-squares),[19] and the additional underlying assumptions are the assumptions of the IV design: exclusion restrictions, monotonicity, and relevance.[52]

The *exclusion restriction* condition requires that any effect of the proposed instrument, in the case of RDD a binary variable indicating the assignment, on the outcome is exclusively through its potential effect on exposure. This assumption is untestable and subject-matter knowledge on the assignment mechanism must be applied to rule out the possibility of any direct effect of the assignment rule on the outcome of interest. This assumption, however, often holds in the RDD settings.

The *monotonicity or "no defiers" assumption* implies a monotonic relationship between the binary variable indicating the assignment, and exposure. In other words, the assignment rule must not increase the exposure in some people and decrease in others, but in every individual exposure assignment must either leave the exposure status unchanged or change it in the same direction of the assignment. Although this assumption is also untestable, its plausibility should be investigated by common sense and observed data patterns.[53]

The *relevance assumption* requires the assignment binary variable (instrument) to be associated with the exposure, and it is empirically verifiable. In the IV setting with the two-stage least squares estimator, it comprises the first stage estimates, i.e., the predicted value of exposure based on the instrument. The same applies to the fuzzy RDD where it is equally important to verify that a discontinuity exists in the relationship between the binary assignment variable and the exposure variable. As with an IV, the stronger the first stage, that is, the larger the discontinuity at the cut-off is, the more efficient and less prone to bias the estimates from the second stage are. The assumption can be tested graphically (*Figure 2*) or formally using, for example, F-statistics. As a rule of thumb, the instrument is declared weak if the F-statistic is less than 10.[54]

### 5. Other sensitivity analyses and validity checks in RDD

The main RDD analysis consists of visually depicting any discontinuity in the outcome of interest by plotting the distribution of the outcome at the two sides of the assignment variable cut-off (Figure 6, simulated data). Any jump in the distribution of the outcome at the assignment variable cut-off can be estimated either using local polynomials,[20,22–24] or by other techniques of the local randomization approach.[21,27]

To strengthen the validity of RDD several additional checks are strongly advised and are summarized below.

(Figure 6 here)

*Discontinuities in average outcomes at values other than the assignment variable cut-off*

One of the most important robustness checks in RDD is the comparison of the main effects for true and artificial (placebo) cut-offs in the assignment variable. Any discontinuity in artificially imposed cut-offs, if unjustified by the knowledge on the underlying mechanisms, is an indication of potentially invalid RDD. This can be done empirically by replacing the true cut-off value by different values of the assignment variable where exposure does not change, and by repeating both the graphical and estimation analysis using standard approaches.

*Sensitivity to bandwidth choice*

The most frequently used RDD estimation methods are non-parametric or local methods that consider only observations in a selected window around the cut-off. Optimal bandwidth size can be selected either ad hoc using common sense and previous knowledge or by data-driven algorithms.[55] In practice, the bandwidth size depends on data availability around the cut-off. Ideally, one would like to use a very narrow window around the cut-off, but this comes at the cost of less precise estimates and lower external validity in most of the RDD settings.[56] Sensitivity analysis with alternative specifications of bandwidth size to check the robustness of the estimated effects is a standard in RDD.

*Sensitivity to observations near the cut-off*

Even if there is no evidence of manipulation in the assignment variable, the observations very near the cut-off are likely to be the most influential when fitting local polynomials. The sensitivity test is also called "donut hole" approach and consists of repeating the analysis on different subsamples where observations are removed in a symmetric distance around the cutoff, starting with those closest to the cut-off and then increasing the distance around cut-off in the attempt to understand the sensitivity of the results to those observations.[57,58]

*Specification of the response function*

Model misspecification is an issue in any analysis, and RDD effects are unbiased only if the functional form of the relationship between the assignment variable and the outcome variable is correctly modeled. One of the frequent issues in RDD is a nonlinear relationship between the assignment variable and the outcome that can be misinterpreted as a discontinuity. Although linear regressions are normally employed in RDD, when the true functional form is unknown it is recommended to include alternative specifications, for example higher order polynomials, in the regression models,[59] and to check the robustness of the effect estimates to multiple specifications.[11] The goodness-of-fit tests can be performed assess model misspecifications. However, it has been shown that higher-order polynomials can lead to overfitting and bias,[60] and it is generally recommended to fit local linear regressions with linear and quadratic forms only, or alternatively to use a local linear nonparametric regression.[19]

## 6. RDD applications in perinatal and pediatric epidemiology

Three review articles so far evaluated the application of RDD in healthcare research.[12–14] The most recent and the only systematic review,[12] that performed searches of articles published until March 2019 in several economic, social, and medical databases, identified 325 studies using RDD in the context of healthcare research. The authors showed an increasing popularity of this design in the past 10 years with the vast majority of studies being applied in the context of specific policies, social programs, health insurance, and education. The review nicely summarizes the mostly used assignment variables (e.g., age, date, socio-economic, clinical, and environmental measures) and the cut-off rules (program eligibility, legislation cut-offs, date of sudden events, and clinical decision-making rules).[12]

From the studies provided in the systematic review[12] we identified studies conducted in the context of perinatal, childhood and adolescent epidemiology with the aim to understand the potential of promoting the use of RDD in the existing birth cohort consortia. We also updated the search until August 13, 2022, using PubMed, with a more specific search strategy detailed in **Supplementary Table 1**.

Of 325 studies from the previous systematic review,[12] we considered 108 studies potentially relevant for perinatal and childhood epidemiology (**Supplementary Table 2**). The additional Pubmed search identified 92 studies, of which 60 were considered relevant for the current review (**Supplementary Table 3**). Numerous studies published in the past three and a half years confirm the increasing popularity of RDD in perinatal and pediatric epidemiology. Although many of the studies from the previous review and the updated search were potentially relevant for childhood health, we considered only studies that were explicitly conducted in the context of interest or that used outcomes assessed in these age groups.

More than 60% of the identified studies (104/168) were conducted in the context of education (educational reforms on schooling initiation and duration), social and welfare programs and policies (e.g., conditional cash transfers, child supplements, parental leaves), and healthcare organization and insurance. About 15% of the studies (25/168) evaluated the effect of specific preventive programs and policies including vaccination campaigns. The remaining studies evaluated clinically relevant research questions (N=18), shock events, social and environmental factors (N=17), changes in guidelines (N=3), and methods relevant for epidemiology (N=1). Interestingly, about 40% of studies published in the past three and a half years focused on clinical settings and shock events, the thematic fields which were quite neglected in the past RDD studies on perinatal and pediatric epidemiology. This is also due to the recent COVID-19 pandemic, which was the focus of several recent RDD studies.[61–65]

Despite an increasing number of RDD studies in the field of perinatal and childhood epidemiology, most of the studies were setting-specific evaluating the effect of specific programs, policies, and sudden events, and are, thus, difficult to implement or replicate using data of birth cohorts. However, there are some previous applications that used data that are typically collected in birth cohorts and may serve as motivating examples for future studies. *Table 2* summarizes some of the assignment variables used for identification of discontinuities in perinatal epidemiology that could be replicated or extended in future birth cohort research.

(Table 2 here)

## 7. Advantages and limitations of the RDD in perinatal and pediatric epidemiology

The RDD is a study design that gained an increasing popularity in health research due to many advantages over other non-experimental study designs. Its estimates and validity checks can be easily presented using simple graphical representations that improve transparency and integrity of the results. The interpretation of the results is intuitive and straightforward, and its validity and the underlying assumptions are relatively weak compared to other study designs and analytical approaches, and many of them can be tested empirically. The RDD circumvents ethical issues of random assignment, and, if the underlying assumptions are met and credible, it can be almost as good as a randomized experiment in measuring treatment effect. To date, most of the RDD studies have primarily used linear regression models for continuous outcomes, but its application is also generalizable to binary, time-to-event, and count outcomes.[66]

Most of the previous perinatal and pediatric epidemiology RDD studies were conducted on the data recovered from registries and administrative databases, which often lack important details and individual-level data. The existing birth cohorts collect a plenty of a very detailed data on parental characteristics, pregnancy outcomes, newborn, infant and later childhood long-term outcomes that have been rarely exploited using RDD. There are several reasons for this, including some of the main limitations of the RDD.

While the RDD has strong internal validity, its external validity is often considered the main caveat. The RD estimate of the treatment effect is limited to the subpopulation of individuals at the discontinuity cut-off and is uninformative about the effect anywhere else. In the sharp RDD the treatment effect is interpreted as the average treatment effect at the cut-off, and it can be generalized and approximated to the average treatment effect only with certain additional assumptions.[67] In the fuzzy RDD the local average treatment effect is estimated at the cut-off, and it is even less generalizable because it is inferred only to the compliers, i.e., the subpopulation of individuals who comply with the assignment rule at the cut-off. In addition to the nature of the estimated effect, the external validity of RDD is further threatened by often setting-specific research questions that cannot be extrapolated and replicated in different populations (e.g., country-specific policies). The utility of RDD also depends on the practical and clinical relevance of the cut-off being studied.

The estimation in RDD implies that we need adequate power for estimating regression line on both sides of the cut-off, i.e., a lot of observations near the cut-off. It has been shown that RDD needs up to three times as many participants as a randomized experiment to have the equivalent power.[68,69] The power will be additionally reduced by the selection of a relatively narrow bandwidth around the cut-off, which is needed to maintain the local randomization, and by small discontinuities in the treatment probability in the case of fuzzy RDD. Although RDD requires a large sample size and it is rarely feasible in a single cohort analysis, it still can be easily powered in the existing large birth cohort consortia.

Additional issue in RDD is the postential contamination of the main exposure of interest by other exposures or interventions that act concomitantly. For example, in the context of perinatal epidemiology, studies assessing the type of neonatal care for very-low-birth-weight newborns (birthweight cut-off <1500 grams) face the problem of distinguishing the effect of neonatal intensive care unit, use of medications (e.g., antibiotics), absence of rooming-in and continuous maternal presence, later initiation of breastfeeding, and other co-existing factors. Finally, despite an increasing popularity of RDD in perinatal and pediatric epidemiology, settings other that policy and program evaluations are still rare. This is probably the most relevant motivation behind the lack of studies within the existing birth cohorts. Still, the assignment variables applied in previous studies can be extended, if appropriate, to additional exposures and used with the same exposures to address different research questions.

## 8. Conclusions

The regression discontinuity design is a powerful approach for causal inference in perinatal and pediatric epidemiology that has several advantages over other non-experimental study designs, including strong internal validity and a relatively weak and testable assumptions. Its widespread use has been hampered by the limited external validity and the rarity of settings outside of program and policy evaluation. The identification of discontinuities and RDD principles should be introduced to researchers who should exploit the utilities of this design in the existing birth cohorts whenever the setting and research question allow.

**Table 1.** The main conditions and assumptions of RDD.

| | Description | Empirical testing |
|---|---|---|
| **Assignment rule condition** | A continuous pre-exposure variable with a clearly defined cut-off value for the exposure assignment | Yes |
| | The same cut-off value is not used to assign the individuals to other exposures | No (theoretical only) |
| | Lack of discontinuities other than the one at the cut-off | Yes |
| **Lack of manipulation in the assignment variable** | The cut-off value is exogenous – unrelated to individuals' value of the assignment variable, and individuals' assignment variable values are not determined by the cut-off of the assignment variable. | Yes (indirectly) |
| **Exchangeability around the assignment variable cut-off** | Similarity of the individuals close to the cut-off in observed characteristics | Yes |
| | Similarity of the individuals close to the cut-off in unobserved characteristics | No |
| | Outcome probability is continuous at the cut-off in the absence of exposure | Yes |

NOTE: The estimates of the causal effect under the fuzzy design require additional assumptions of (i) the relevance of the assignment binary variable which needs to be associated with the exposure, (ii) the monotonicity of the treatment selection response to the assignment variable at the cut-off, and (iii) the local exclusion restriction assumption.

**Table 2**. Assignment variables and exposures as possible RDD models for birth cohort research.

| Assignment variable | Determined cut-off values | Possible exposures / interventions |
|---|---|---|
| **Birth weight** | Low-birth weight (<2500 grams) Very low birth weight (<1500 grams) Extremely low birth weight (<1000 grams) High birth weight (>4000 or >5000 grams) | Extra neonatal care Neonatal intensive care unit Rooming-in and mother-child bonding Breastfeeding Caesarean section Specific treatments (e.g., probiotic supplementation, surfactant therapy) Setting-specific health insurance and supplemental benefits |
| **Gestational age** | Preterm (<37 gestational weeks) Very preterm (<32 gestational weeks) Extremely preterm (<28 gestational weeks) | |
| **Maternal age at conception** | <18 years >35 years >40 years | Minimum cigarette/alcohol purchase age Screening and procedures for high-risk pregnancies |
| **Socioeconomic measures** | Setting-specific | Social, welfare, and cash transfer programs |
| **Parity** | Setting-specific | Health insurance policies |
| **Age or date/year of birth** | Setting-specific | Introduction of: <ul><li>Vaccination campaigns</li><li>Pregnancy-specific guidelines</li><li>Maternity/paternity leave policies</li><li>Child-support grants</li><li>Social and welfare programs</li></ul> |
| **Calendar time** | Setting-specific | Introduction of specific programs and policies Shock events |
| **Clinical measures** | Setting-specific | Treatment initiation Preventive programs |
| **Environmental measures** | Setting-specific | Local interventions (e.g., to reduce air-pollution) |

**Figure 1.** Causal directed acyclic graph (DAG) of the regression discontinuity design. *X* represents the assignment variable which cut-off value *c* determines the eligibility criteria (*D*) for an exposure / treatment (*E*). Y denotes an outcome, while U represents measured or unmeasured confounding factors. Panel A: DAG depicting a causal model underlying a regression discontinuity design. Panel B: Causal graph for the regression discontinuity design in the close vicinity around the assignment variable cut-off. Note that if the assignment rule is deterministic (i.e. sharp regression discontinuity design) *D* is equal to *E*.

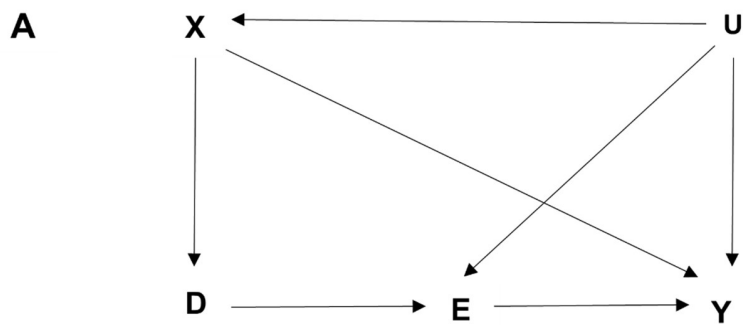

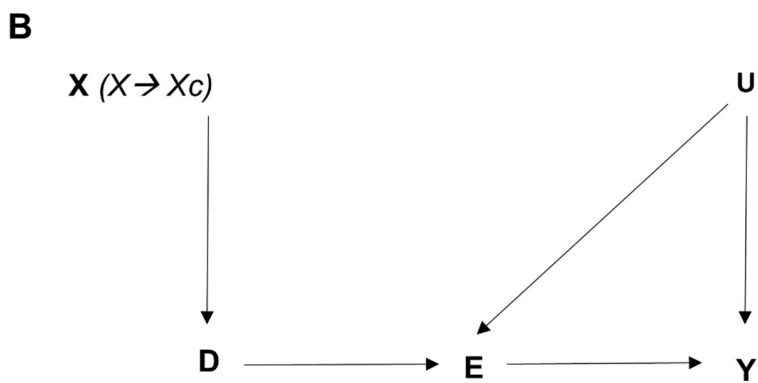

**Figure 2.** Hypothetical sharp and fuzzy regression discontinuity design. Panel A: Sharp (deterministic) regression discontinuity design. Panel B: Fuzzy (probabilistic) regression discontinuity design. Simulated data

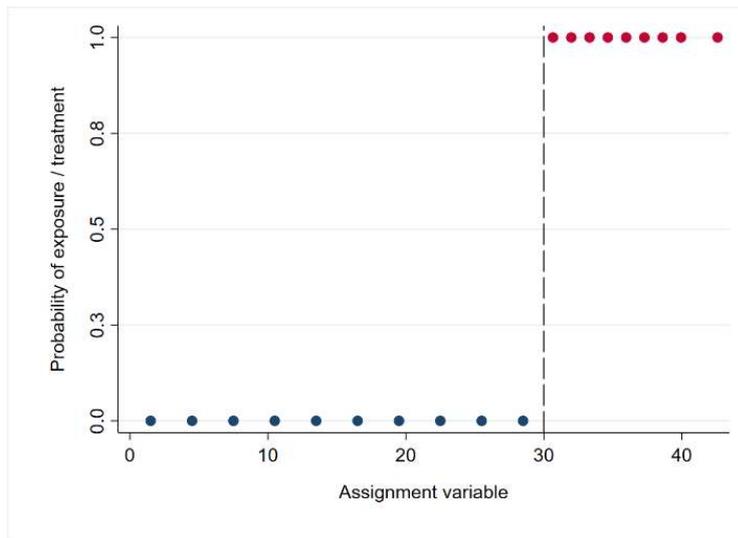

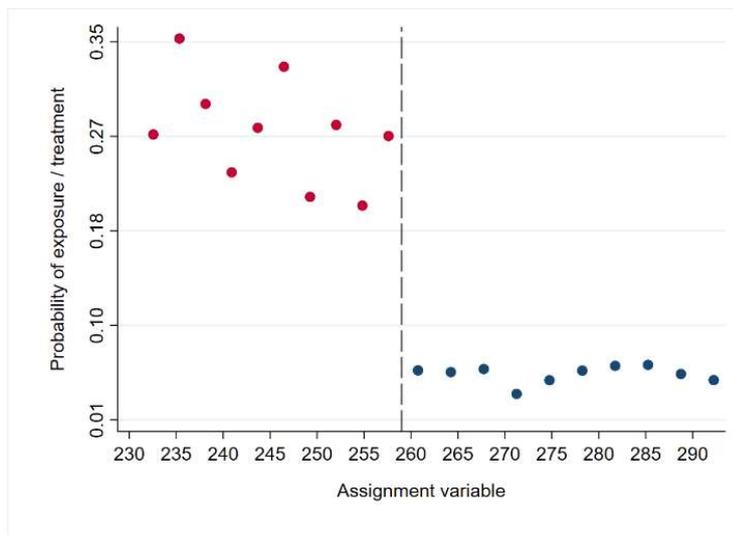

**Figure 3.** The probability of exposure as a function of assignment variable with different bin widths for graphical presentations. Panel A: 20 bins. Panel B: 40 bins. Panel C: 100 bins. Simulated data.

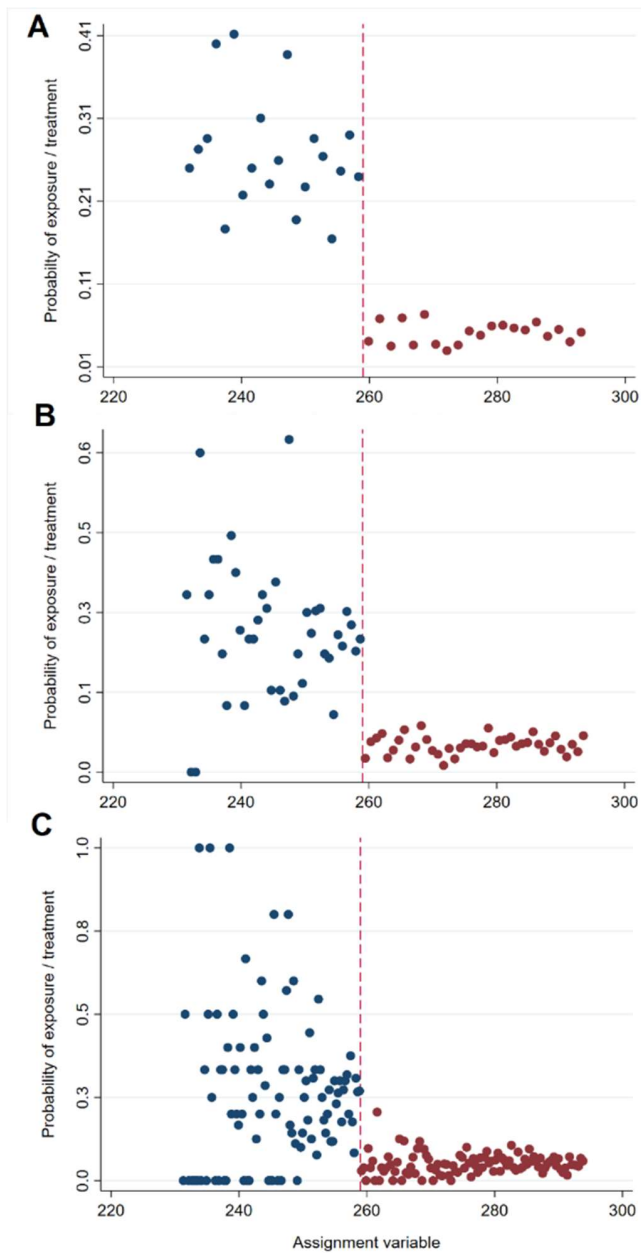

**Figure 4.** Graphical representation of the assignment variable manipulation**.** Panel A: No evidence of manipulation in assignment variable at cut-off. Panel B: Evidence of manipulation in assignment variable at cut-ff. Note that the graph represents the centered assignment variable. Simulated data

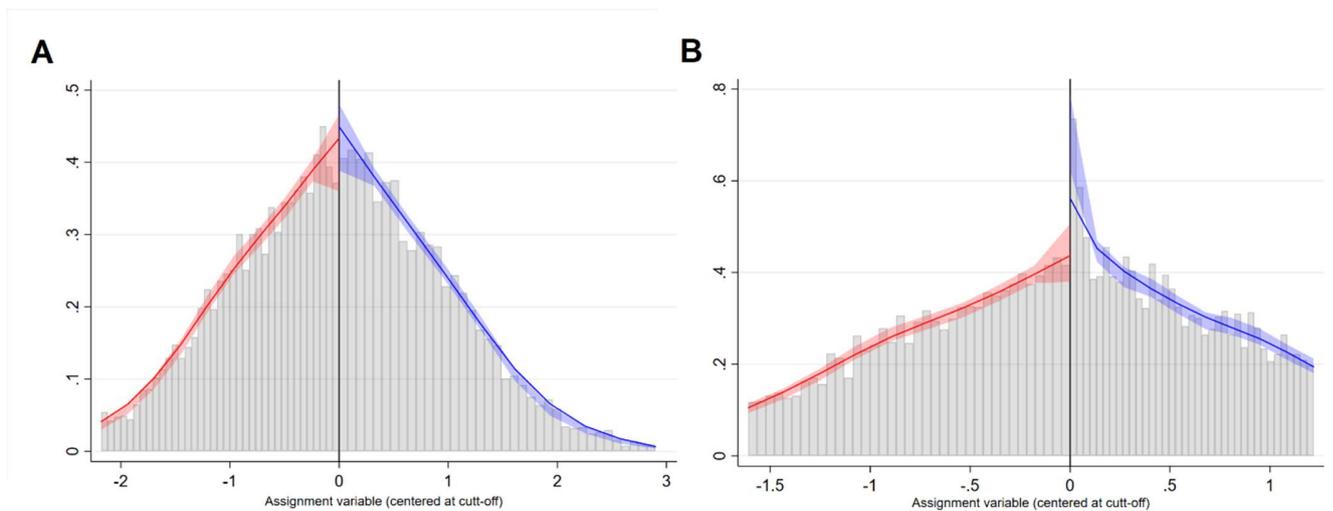

**Figure 5.** Graphical representation of RD for predetermined covariates for a hypothetical example of birth weight as an assignment variable for an intensive neonatal care. Data from the NINFEA birth cohort.

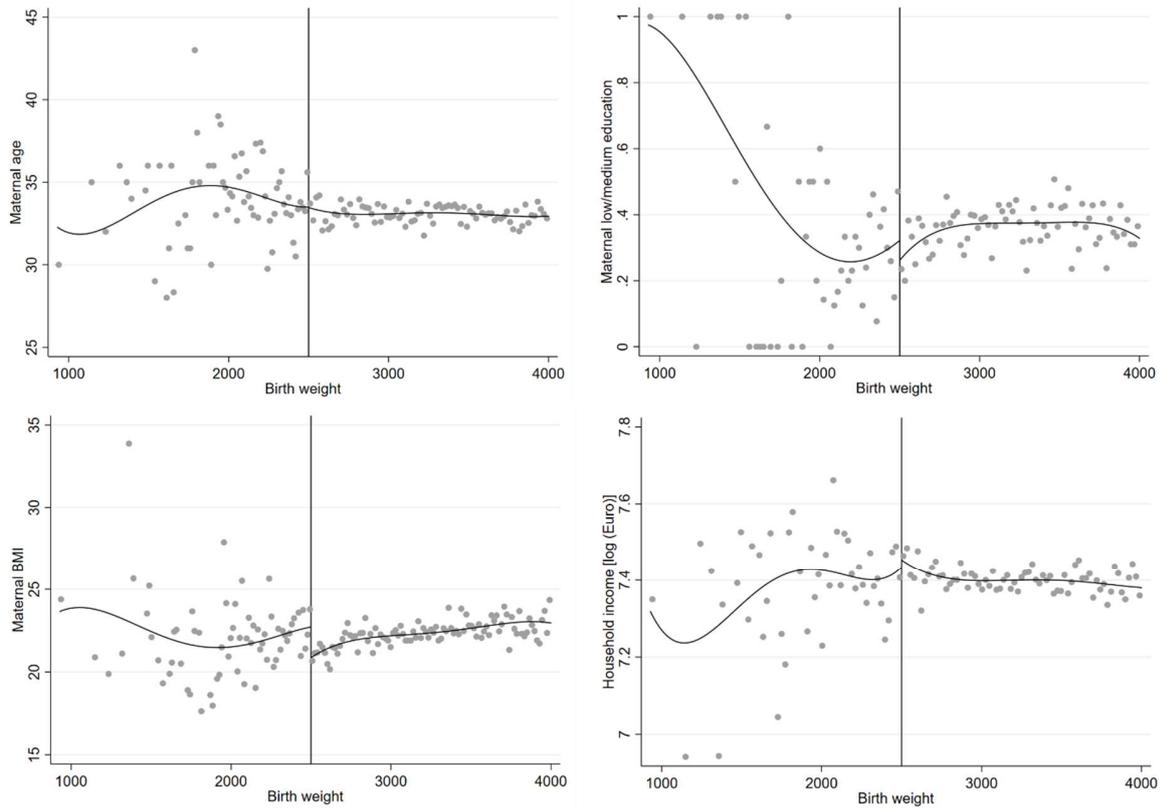

**Figure 6.** Discontinuity in the outcome (weight at 18 months of age) at the assignment variable cut-off (birth weight). Simulated data.

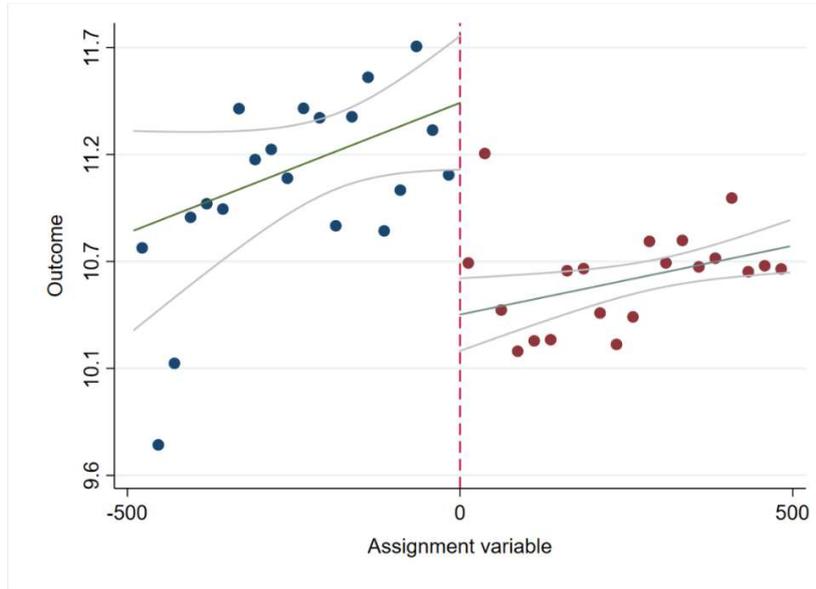

# Supplementary Material

**Supplementary Table 1. PubMed search strategy**

| Database | Date | Search strategy | Filter | Number of results |
|---|---|---|---|---|
| PubMed | 13/08/2022 | ("regression discontinuity" OR "regression discontinuity design") AND (child* OR offspring OR newborn OR infan* OR neonatal OR perinatal OR adolescen* OR teenag* OR mater* OR mother OR pater* OR father OR parent* OR pregnan* OR birth) | From 2019 | 92 |
| | | (("regression discontinuity"[All Fields] OR "regression discontinuity design"[All Fields]) AND ("child*"[All Fields] OR ("offspring"[All Fields] OR "offspring s"[All Fields] OR "offsprings"[All Fields]) OR ("infant, newborn"[MeSH Terms] OR ("infant"[All Fields] AND "newborn"[All Fields]) OR "newborn infant"[All Fields] OR "newborn"[All Fields] OR "newborns"[All Fields] OR "newborn s"[All Fields]) OR "infan*"[All Fields] OR ("infant, newborn"[MeSH Terms] OR ("infant"[All Fields] AND "newborn"[All Fields]) OR "newborn infant"[All Fields] OR "neonatal"[All Fields] OR "neonate"[All Fields] OR "neonates"[All Fields] OR "neonatality"[All Fields] OR "neonatals"[All Fields] OR "neonate s"[All Fields]) OR ("perinatal"[All Fields] OR "perinatally"[All Fields] OR "perinatals"[All Fields]) OR "adolescen*"[All Fields] OR "teenag*"[All Fields] OR "mater*"[All Fields] OR ("mother s"[All Fields] OR "mothered"[All Fields] OR "mothers"[MeSH Terms] OR "mothers"[All Fields] OR "mother"[All Fields] OR "mothering"[All Fields]) OR "pater*"[All Fields] OR ("father s"[All Fields] OR "fathered"[All Fields] OR "fathers"[MeSH Terms] OR "fathers"[All Fields] OR "father"[All Fields] OR "fathering"[All Fields]) OR "parent*"[All Fields] OR "pregnan*"[All Fields] OR ("birth s"[All Fields] OR "birthed"[All Fields] OR "birthing"[All Fields] OR "parturition"[MeSH Terms] OR "parturition"[All Fields] OR "birth"[All Fields] OR "births"[All Fields]))) AND (2019:2022[pdat]) | | |
| | | ***Translations*** | | |
| | | offspring: "offspring"[All Fields] OR "offspring's"[All Fields] OR "offsprings"[All Fields] | | |
| | | newborn: "infant, newborn"[MeSH Terms] OR ("infant"[All Fields] AND "newborn"[All Fields]) OR "newborn infant"[All Fields] OR "newborn"[All Fields] OR "newborns"[All Fields] OR "newborn's"[All Fields] | | |
| | | neonatal: "infant, newborn"[MeSH Terms] OR ("infant"[All Fields] AND "newborn"[All Fields]) OR "newborn infant"[All Fields] OR "neonatal"[All Fields] OR "neonate"[All Fields] OR "neonates"[All Fields] OR "neonatality"[All Fields] OR "neonatals"[All Fields] OR "neonate's"[All Fields] | | |
| | | perinatal: "perinatal"[All Fields] OR "perinatally"[All Fields] OR "perinatals"[All Fields] | | |
| | | mother: "mother's"[All Fields] OR "mothered"[All Fields] OR "mothers"[MeSH Terms] OR "mothers"[All Fields] OR "mother"[All Fields] OR "mothering"[All Fields] | | |
| | | father: "father's"[All Fields] OR "fathered"[All Fields] OR "fathers"[MeSH Terms] OR "fathers"[All Fields] OR "father"[All Fields] OR "fathering"[All Fields] | | |
| | | birth: "birth's"[All Fields] OR "birthed"[All Fields] OR "birthing"[All Fields] OR "parturition"[MeSH Terms] OR "parturition"[All Fields] OR "birth"[All Fields] OR "births"[All Fields] | | |

**Supplementary Table 2. Regression discontinuity studies focused on perinatal, childhood and adolescent epidemiology identified from a previous systematic review on the use of RDD in health research (*Hilton Boon Epidemiology 2021, updated until March 2019*)**

| AUTHOR | SETTING | EXPOSURE/INTERVENTION | ASSIGNMENT VARIABLE | Outcome |
|---|---|---|---|---|
| Almond (2010)[1] | Clinical | Neonatal intensive care | Birthweight | Infant mortality |
| Bharadwaj (2013)[2] | Clinical | Neonatal intensive care - extra medical attention and lung surfactant therapy | Birthweight | Child cognitive development - academic achievement, mortality |
| Daysal (2013)[3] | Clinical | Obstetrician supervision of preterm birth | Gestational age | Seven- and 28-day mortality, Apgar score |
| Belenkiy (2010)[4] | Healthcare/Insurance | Health insurance | Age | Obstetric treatment intensity |
| Garrouste (2011)[5] | Healthcare/Insurance | Reimbursement eligibility | Down syndrome risk score | Amniocentesis and fetal health |
| Almond (2011)[6] | Healthcare/Insurance | Length of hospital stay | Clock time | Maternal and newborn health |
| De La Mata (2012)[7] | Healthcare/Insurance | Medicaid | Family income | Healthcare utilization, health status, obesity, school sickness absence |
| Koch (2013)[8] | Healthcare/Insurance | Public health insurance for children | Family income | Healthcare utilization and expenditure |
| Camacho (2013)[9] | Healthcare/Insurance | Subsidized Regime health insurance for the poor | Poverty index | Neonatal health (birthweight, Apgar score), prenatal care |
| Palmer (2015)[10] | Healthcare/Insurance | Public health insurance for preschool children | Age | Healthcare utilization, expenditure, substitution (crowdout) |
| Koch (2015)[11] | Healthcare/Insurance | Public health insurance | Family income | Parents' self-reported health and preventive care usage |
| Han (2016)[12] | Healthcare/Insurance | Children's Medical Subsidy Program | Age | Healthcare utilization and expenditure |
| Bhowmick (2016)[13] | Healthcare/Insurance | Community health worker programme | Population | Pregnancy and child health outcomes |
| Laughery (2016)[14] | Healthcare/Insurance | Health Professional Shortage Area designation | Number of GPs per 10,000 population | Hospitalizations, mortality, prenatal care, neonatal health |
| Lee (2017)[15] | Healthcare/Insurance | Medicaid plan (fee for service vs managed care) | Birthweight | Hospital readmission, length of stay, mortality |
| Bernal (2017)[16] | Healthcare/Insurance | Seguro Integral de Salud (social health insurance) | Household Targeting Index (welfare index) | Vaccines, birth control |
| Del Bono (2011)[17] | Guidelines | UK Committee on Safety of Medicines health warning on combined oral contraceptives and risk of VTE | Calendar time | Daily average numbers of conceptions, abortions, and live births; neonatal health outcomes |
| Jensen (2015)[18] | Guidelines | Caesarean section for breech births (changes in guidelines) | Calendar time | Apgar score, physician visits, hospital admissions, complications, infections |
| Hamad (2016)[19] | Guidelines | Publication of national guidelines on nutrition during pregnancy | Calendar time | Gestational weight gain |
| Rashad (1992)[20] | Preventive/Vaccination programs | National Control of Diarrheal Diseases Project | Calendar time | Infant mortality |
| Waller (2003)[21] | Preventive/Vaccination programs | Change in cigarette prices | Calendar time | Youth smoking prevalence and mean cigarettes smoked per day |
| Schanzenbach (2009)[22] | Preventive/Vaccination programs | National School Lunch Program | Income to poverty ratio | Child obesity |
| Ziegelhöfer (2012)[23] | Preventive/Vaccination programs | Rural water supply and hygiene education programme | Investment cost per inhabitant | Prevalence of diarrheal disease in children under 5 years |

| Peckham (2012)[24] | Preventive/Vaccination programs | National School Lunch Program | Family income to poverty ratio | Obesity (BMI, waist-to hip ratio, percentage body fat) |
|---|---|---|---|---|
| Meller (2014)[25] | Preventive/Vaccination programs | PANN2000 food supplementation and health check programme | Poverty index | Child mortality, fertility |
| Yan (2014)[26] | Preventive/Vaccination programs | Minimum cigarette purchase age | Maternal age at conception | Prenatal smoking, infant health measures |
| Dykstra (2015)[27] | Preventive/Vaccination programs | Gavi vaccination aid programme | Multiple per capita national income | Vaccination rates and child mortality |
| McMahon (2015)[28] | Preventive/Vaccination programs | School-based food intervention to reduce radionuclide exposure | Calendar time | Blood markers, anemia, respiratory and immune diseases |
| Smith (2015)[29] | Preventive/Vaccination programs | HPV vaccination | Quarter of birth | Composite indicator of sexual behavior |
| Olsho (2015)[30] | Preventive/Vaccination programs | US Dept of Agriculture Fresh Fruit and Vegetable Program | Proportion of students eligible for free or reduced-price meals | 24-hour dietary intake |
| Helleringer (2016)[31] | Preventive/Vaccination programs | Mass vaccination campaign (supplementary immunization activity) | Date of birth | Routine polio vaccination |
| Almond (2016)[32] | Preventive/Vaccination programs | Fitnessgram obesity report cards for schoolchildren | Body mass index (BMI) | BMI and weight |
| Gertner (2016)[33] | Preventive/Vaccination programs | Home visits to promote improved nutritional knowledge and child health | Distance to geographical boundary | Nutritional knowledge and anthropometric measures |
| Bakolis (2016)[34] | Preventive/Vaccination programs | Smoking ban | Calendar time | Birth outcomes |
| Billings (2018)[35] | Preventive/Vaccination programs | Interventions for elevated blood lead levels in children | Blood lead levels | Adolescent Antisocial Behavior Index |
| Pitt (1999)[36] | Social and welfare programs | Group-based credit programmes for the poor | Acres of land owned by household | Contraceptive use and fertility |
| Santos (2006)[37] | Social and welfare programs | BabyFirst home visit programme | Family Stress Checklist score | Family social support, parental mental health, parenting outcomes |
| Ludwig (2007)[38] | Social and welfare programs | Head Start | County-level poverty index | Child mortality |
| Urquieta (2009)[39] | Social and welfare programs | Oportunidades poverty alleviation programme | Poverty index | Skilled attendance at delivery |
| Alam (2011)[40] | Social and welfare programs | Female School Stipend Program (conditional cash transfer) | District literacy rate | Sexual and fertility decisions (early marriage and childbearing |
| Rosero (2011)[41] | Social and welfare programs | Early childhood programmes (home visits and childcare centres for poor families) | Programme proposal quality score | Multiple child health and development measures; maternal stress and depression |
| de Brauw (2011)[42] | Social and welfare programs | Comunidades Solidarias Rurales | Municipal poverty score | Prenatal and postnatal care, skilled attendance, birth at health facility |
| Janssens (2011)[43] | Social and welfare programs | Women's empowerment and health education programme | Age | Child vaccinations |
| Rieck (2012)[44] | Social and welfare programs | Paid paternity leave | Date of birth | Parental sickness absence |
| Medina (2013)[45] | Social and welfare programs | Unemployment Subsidy and retraining | Welfare index | Children's weight, height, BMI, Apgar score |
| Bor (2013)[46] | Social and welfare programs | Extension of eligibility for Child Support Grant | Date of birth | Time to first pregnancy from age 14 (teenage pregnancy) |
| Tibone (2013)[47] | Social and welfare programs | US foreign aid policy change | Month and year of conception | Abortion rates |
| You (2013)[48] | Social and welfare programs | Formal microcredit | Predicted probability of borrowing microcredit | Child malnutrition (BMI, anemia, zinc deficiency) |
| González (2013)[49] | Social and welfare programs | Universal child benefit | Calendar time | Incidence of conceptions and abortions |
| Carneiro (2014)[50] | Social and welfare programs | Head Start | Family income | Health measures from CNLSY longitudinal survey |
| Sun (2014)[51] | Social and welfare programs | Increased women's bargaining power following divorce reform | Month and year of conception | Sex ratio of second children following firstborn girls; birth spacing; child caloric intake; husband's alcohol and cigarette consumption |
| Filmer (2014)[52] | Social and welfare programs | Scholarships for poor children | Dropout risk score | Teenage pregnancy |
| Cogneau (2015)[53] | Social and welfare programs | National boundaries | Distance from border | Children's height-for age, access to safe water |
| Carranza Barona (2015)[54] | Social and welfare programs | Bono de Desarrollo Humano (conditional cash transfer) | Selben welfare index | Exclusive breastfeeding in the first six months of life |
| El-Kogali (2015)[55] | Social and welfare programs | Community development programme | District poverty level | Child growth and nutrition |
| Beuchert (2016)[56] | Social and welfare programs | Maternity leave policy change | Calendar time | Hospital visits |
| Deepti Thomas (2016)[57] | Social and welfare programs | National Rural Employment Guarantee Act | State development index | Child and maternal mortality, vaccinations |
| Sachdeva (2016)[58] | Social and welfare programs | National rural road construction programme | Population | Prenatal care and contraception, healthcare supply |
| Cygan-Rehm (2016)[59] | Social and welfare programs | Parental benefit based on net earnings | Calendar time | Fertility and birth spacing |
| You (2016)[60] | Social and welfare programs | Access to microcredit | Propensity to borrow from rural microcredit schemes | Parental report of child health |
| Moreno (2017)[61] | Social and welfare programs | Human Development Bonus (conditional cash transfer) | Household poverty index | Child chronic stunting |
| Cattaneo (2017)[62] | Social and welfare programs | Head Start | County-level poverty index | Child mortality |
| Guertzgen (2018)[63] | Social and welfare programs | Reform of maternity leave legislation | Month of birth | Long-term sickness |
| Tang (2017)[64] | Social and welfare programs | Head Start | Calendar time | Child cognitive development and behaviors |
| Guldi, (2018)[65] | Social and welfare programs | Supplemental Security Income benefit | Birthweight | Infant mortality, child motor skill development, parenting behaviors, inequalities |
| Rahman (2018)[66] | Social and welfare programs | Safe motherhood scheme (conditional cash transfer) | Parity | Use of maternal and child health services |
| Deutscher (2018)[67] | Social and welfare programs | Birth shifting in response to baby bonus policy | Calendar time | Birthweight, gestation length, child developmental score |
| Garcia Hombrados (2018)[68] | Social and welfare programs | Increased legal age of marriage for women under the Revised Family Code | Age | Infant mortality |
| Gormley (2005)[69] | Education | Universal prekindergarten program | Birthdate | School readiness |
| Wong (2008)[70] | Education | State prekindergarten programme | Birthdate | Children's cognitive skills/school readiness |
| Monstad (2008)[71] | Education | Reform that increased years of compulsory schooling | Age relative to year of reform | Number of children and maternal age at first birth |
| Coburn (2009)[72] | Education | Prekindergarten programme | Age | Brigance Screen age equivalent scores |
| Lindeboom (2009)[73] | Education | Additional year of schooling | Year of birth | Child height, weight, morbidity; parental BMI, chronic disease, fertility |
| Zhang (2009)[74] | Education | Years of formal schooling | Age | Children's bodyweight, fruit and vegetable consumption |

| Elder (2010)[75] | Education | School starting age | Date of birth | ADHD symptoms, diagnosis and treatment |
|---|---|---|---|---|
| Evans (2010)[76] | Education | School starting age | Date of birth | ADHD diagnosis and treatment |
| Dickert-Conlin (2010)[77] | Education | State cutoff dates for school eligibility | Calendar time | Birth timing |
| Lipsey (2011)[78] | Education | Voluntary pre-kindergarten programme | Birthdate | School readiness |
| McCrary (2011)[79] | Education | School starting age | Date of birth | Fertility, birthweight, prematurity |
| Anderson (2011)[80] | Education | Years of early primary education | Date of birth | Children's BMI |
| Nakamura (2012)[81] | Education | Maternal schooling | Month and year of birth | Children's weight, fruit and vegetable consumption, exercise |
| Greenwood (2012)[82] | Education | College opening | Calendar time | Births to teenage mothers |
| Weiland (2013)[83] | Education | Public Schools prekindergarten programme | Date of birth | Cognitive, executive function and emotional development |
| Jakobsson (2013)[84] | Education | Class size | Class size | Adolescent mental health and wellbeing measures |
| Ankara (2015)[85] | Education | Extension of compulsory schooling | Date of birth | Fertility and child mortality |
| Chen (2015)[86] | Education | School starting age | Age | Inattentive/hyperactive behavior |
| Grépin (2015)[87] | Education | 1980 School Reform in Zimbabwe | Age | Child mortality, vaccinations, antenatal care |
| Samarakoon (2015)[88] | Education | Education (longer school year in 1978) | Year of birth | Fertility and reproductive health behaviors |
| Makate (2016)[89] | Education | Universal Primary Education policy | Age | Neonatal and child mortality |
| Schwandt (2016)[90] | Education | School starting age | Age | ADHD prevalence and medications |
| Tan (2017)[91] | Education | School entry cutoff date | Date of birth | Motherhood by age 17 and age 20 |
| Weitzman (2017)[92] | Education | Extension to compulsory schooling by 5 years | Age | Maternal health: complications during and after pregnancy, use of contraception |
| Adu Boahen (2018)[93] | Education | Free compulsory universal primary education | Year of birth | Sexual behavior and fertility |
| Ali (2018)[94] | Education | Increased years in primary education | Date of birth | Fertility |
| Ali (2018)[95] | Education | Parental education | Date of birth | Child mortality |
| Kan (2018)[96] | Education | Increased compulsory education | Date of birth | Fertility |
| Dee (2018)[97] | Education | Delay starting school by one year | Age | Child mental health: total difficulties, emotional, conduct, hyperactivity, peer problems, prosocial behavior |
| Makate (2018)[98] | Education | 1980 School Reform in Zimbabwe | Calendar year | Child height-for-age and weight-for-age |
| Makate (2018)[99] | Education | Universal primary education programme | Age | Teenage childbirth |
| Keats (2018)[100] | Education | Universal primary education programme | Year of birth | Fertility, child mortality, vaccinations, malnutrition |
| Chay (2003)[101] | Shock events, environmental and social factors | Clean Air Act Amendments (1970) | Total suspended particulates (air pollution regulatory threshold) | Infant mortality |
| Yang (2008)[102] | Shock events, environmental and social factors | Clean Air Act Amendments (1970) | Total suspended particulates (air pollution regulatory threshold) | Infant mortality |
| Dell (2010)[103] | Shock events, environmental and social factors | The mita - a forced labour system in operation 1573-1812 | Latitude and longitude | Stunted growth in children |
| Huang (2013)[104] | Shock events, environmental and social factors | Great Famine 1959-61 | Year of birth | Cognitive functioning |

| Sotomayor (2013)[105] | Shock events, environmental and social factors | In-utero exposure to natural disasters | Year of birth | Hypertension, diabetes, high cholesterol in adulthood |
|---|---|---|---|---|
| Bhalotra (2014)[106] | Shock events, environmental and social factors | Rise in share of elected officials who are Muslim | Vote margin in close elections | Neonatal and infant mortality |
| Zhong (2014)[107] | Shock events, environmental and social factors | Number of siblings | Year of birth | Child health (height, self-assessed health, BMI) |
| Sanders (2015)[108] | Shock events, environmental and social factors | Clean Air Act Amendments (1970) | Total suspended particulates (air pollution regulatory threshold) | Sex ratio of live births as estimate of averted fetal losses |

**Supplementary Table 3. Regression discontinuity studies focused on perinatal, childhood and adolescent epidemiology identified from a PubMed search (published from January 1, 2019, until August 13, 2022)**

| AUTHOR | SETTING | EXPOSURE/INTERVENTION | ASSIGNMENT VARIABLE | Outcome |
|---|---|---|---|---|
| Daysal (2019)[109] | Clinical | Obstetrician supervision of preterm birth | Gestational age | Mortality, fetal distress, emergency C-section |
| Hutcheon (2020)[110] | Clinical | Antenatal corticosteroid administration | Gestational age | Early child development score |
| Brilli (2020)[111] | Clinical | Neonatal care for high birthweight newborns | Birthweight | Neonatal intensive care, use of antibiotics, infant mortality |
| Bommer (2020)[112] | Clinical | Routine probiotics supplementation | Gestational age | Anthropometric development, late-onset sepsis among moderately preterm newborns |
| Song (2020)[113] | Clinical | Tight glycemic control in pregnant women diagnosed with GDM | Composite score based on oral glucose tolerance tests | Maternal and neonatal outcomes |
| Tymejczyk (2020)[114] | Clinical | Introduction of pediatric or general Treat All program: Universal Antiretroviral Treatment | Calendar time | Rapid treatment initiation among young adolescents with HIV |
| Worsham (2021)[115] | Clinical | Opioid prescription in adolescents | Age | Opioid-related adverse events |
| Chyn (2021)[116] | Clinical | Early-life interventions for low-birth-weight newborns | Birthweight | Supplemental security income program participation, grade repetition, special education services, test scores, high school disciplinary offenses, college preparation, maternal care and stress, maternal labor market activity and fertility, child medicaid enrollment and expenditures by age 2 |
| Brazier (2021)[117] | Clinical | Introduction of pediatric or general Treat All program: Universal Antiretroviral Treatment | Calendar time | CD4 testing and monitoring of patients with HIV: adults, adolescents and children |
| Holzhausen (2021)[118] | Clinical | Difference between objective and subjective evaluation of sleep duration among children | Age | Sleep duration |
| Furzer (2022)[119] | Clinical | Discrepancy between teacher and parent ADHD assessments | Date of birth | Over-diagnosis or under-diagnosis of ADHD |
| Kim (2022)[120] | Clinical | Overweight/obese classification | Children BMI at earlier age | Children BMI at later age |
| Hutcheon (2022)[121] | Clinical | Antenatal corticosteroid administration | Gestational age | Child ADHD medication data |
| Tennant (2022)[122] | Clinical | Clinical diagnosis of gestational diabetes | Fasting plasma glucose | Birthweight, large for gestational age, caesarean section, shoulder dystocia, perinatal death |
| van der Linde (2022)[123] | Clinical | Treatment in a specialized pediatric hemato-oncology vs. adult hemato-oncology setting | Age | 5-Year Survival in Acute Lymphoblastic Leukemia |
| Johansson (2019)[124] | Healthcare/Insurance | Primary care co-payments | Age | Visits to primary care physician among young adults (18–22 years) |
| Nishioka (2021)[125] | Healthcare/Insurance | Medical expenditure after marginal cut of cash benefit | Child age | Household healthcare costs, outpatient visits |
| Liu (2021)[126] | Healthcare/Insurance | Medicaid managed care versus Medicaid fee-for-service | Birthweight | Emergency department use and hospitalization during the first 6 and 12 |
| | | | | months of life among low-birth-weight infants |
| Di Giacomo (2022)[127] | Healthcare/Insurance | Policy that eliminated co-payments for noninvasive screening tests | Calendar time | Take-up of prenatal tests |
| Benny (2019)[128] | Preventive/Vaccination programs | Minimum legal drinking age laws | Age | Police-reported violent victimization events among youth aged 14-22 years |
| Raifman (2020)[129] | Preventive/Vaccination programs | Minimum purchaser age for the sale of handguns | Age | Adolescent suicide rate |
| Musarandega (2020)[130] | Preventive/Vaccination programs | Interventions within the Elizabeth Glaser Paediatric AIDS Foundation consortium and the Children's Investment Fund Foundation | Calendar time | Critical Path Indicators (mother-to-child transmission of HIV) |
| Hategeka (2020)[131] | Preventive/Vaccination programs | Wide-scale HPV vaccination programme | Calendar time | Sexual behaviors among adolescent women |
| Frio (2021)[132] | Preventive/Vaccination programs | HPV vaccination campaign | Age | Risky sexual behavior in girls |
| Hirani (2021)[133] | Preventive/Vaccination programs | Vaccination reminder letter policy | Date of birth | Vaccination adherence |
| Motta (2021)[134] | Preventive/Vaccination programs | Effect of Wakefield et al. (1998) on skepticism about MMR vaccine | Date of study publication | Monthly MMR injury claims as a proxy of vaccine safety |
| de Chaisemartin (2021)[135] | Preventive/Vaccination programs | Newborns' BCG vaccination | Birth year | COVID-19 cases and hospitalizations |
| Ahammer (2022)[136] | Preventive/Vaccination programs | Minimum legal drinking ages | Age | Teenage drinking behavior and morbidity |
| Higgerson (2019)[137] | Social and welfare programs | Free access to swimming pools | Age | Children's participation in swimming |
| Velasco (2020)[138] | Social and welfare programs | Conditional cash transfer | Socioeconomic status index | Contraceptive behavior among women of childbearing age |
| Chuard (2020)[139] | Social and welfare programs | Policy reform on the duration of paid parental leave: the share of mothers who work up to the 32nd week of pregnancy | Calendar time | Birthweight, gestational length, Apgar scores |
| Brauw (2020)[140] | Social and welfare programs | Conditional cash transfer | Distance to cluster based on municipality level poverty rate and the severe stunting rate among first graders | Maternal health service utilization |
| Batyra (2021)[141] | Social and welfare programs | Changes in minimum-age-at-marriage laws | Age at law implementation | Teenage marriage |
| Alfaro-Hudak (2022)[142] | Social and welfare programs | The Supplemental Nutrition Assistance Program | Poverty-income ratio | Cardiometabolic risk factors in children and adolescents |
| González (2022)[143] | Social and welfare programs | Universal child benefit (cash transfer) | Date of birth | Birth weight, still birth, early neonatal death in the first 24 hours, normality of birth, C-section, weeks of gestation. |
| Gao (2022)[144] | Social and welfare programs | Reproductive health policy | Year of birth | Income, years of schooling |
| Xu (2022)[145] | Social and welfare programs | China's New Rural Pension programme | Age of elderly | Physical and mental health of rural children |
| Hong (2019)[146] | Education | Universal pre-kindergarten program | Date of birth | Healthcare utilization and recorded diagnoses (physical health conditions) |
| Courtin (2019)[147] | Education | Education reform that raised the minimum school leaving age | Age | Biomarkers of cardiovascular, metabolic, organ and immune function |
| Plotnikov (2020)[148] | Education | Raising of the school leaving age from 15 to 16 years | Year of birth | Myopia |
| Butler (2020)[149] | Education | Attendance of selective schools | Test scores taken at age 11 | Overall health, mental health, limitation due to health problems, number of chronic diseases, risk of death by age 60 |
| He (2021)[150] | Education | School entry cut-off date | Date of birth | Myopia |
| Bonilla (2021)[151] | Education | Ethnic studies courses in schools | Grade Point Average (GPA) | educational attainment and engagement |
| Chang (2021)[152] | Education | Working mothers' school entry age (school entry policy) | Date of birth | Age at childbirth |
| Grätz (2022)[153] | Education | Increasing the Minimum School-Leaving Age (parents) | Year of birth | Educational attainment of children |
| Amwonya (2022)[154] | Education | Female education (years of schooling) | Year of birth | Maternal health care utilization |
| Zhang (2022)[155] | Education | Compulsory Education Law | Month of birth | Mean spherical equivalent refractive error and uncorrected visual acuity |
| Muchomba (2022)[156] | Education | Change in reform: Years of schooling | Year of birth | Anemia and BMI of reproductive age women |
| Ye (2022)[157] | Education | Compulsory education reforms: mandatory and free primary and lower secondary education | Year and province of birth | Allostatic load in adults |
| Raucher (2022)[158] | Education | Compulsory schooling laws | Year of birth | Migration |
| Gong (2020)[159] | Shock events, environmental and social factors | Rustication program in China | Year of birth | Adult physical and mental health |
| Fang (2020)[160] | Shock events, environmental and social factors | Fetal exposure to famine | Date of birth | Anthropometric measures in adulthood |
| Aso (2020)[161] | Shock events, environmental and social factors | Fukushima Daiichi Nuclear Power Plant accident | Calendar time | Use of Computed Tomography in children with mild head injuries |
| Been (2020)[162] | Shock events, environmental and social factors | COVID-19 mitigation measures | Calendar time | Incidence of preterm birth |
| Buitrago (2021)[163] | Shock events, environmental and social factors | Cease-fire declared during the Colombian peace (exposure to conflict events during pregnancy) | Date of conception | Pregnancy outcomes (miscarriage, stillbirth, perinatal mortality) |
| Bakolis (2021)[164] | Shock events, environmental and social factors | Lifting of COVID-19 'lockdown' policy | Calendar time | Mental health service use and mortality (including children and adolescents) |
| Takaku (2021)[165] | Shock events, environmental and social factors | COVID-19 school closure | Age | Child outcomes, domestic violence, satisfaction with marriage, risk of divorce, quality of marriage |
| Hinderslay (2022)[166] | Shock events, environmental and social factors | COVID-19 lockdown | Calendar time | Domestic violence and abuse, mental illness during pregnancy |
| Thomas (2022)[167] | Shock events, environmental and social factors | COVID-19 pandemic (within the ongoing social programme) | Programme visit number | Parenting/caregiver behavior |
| Maddow-Zimet (2020)[168] | Epidemiological methods | Changes in response options on reported pregnancy intentions within the Pregnancy Risk Assessment Monitoring System | Calendar time | Pregnancy intentions |